# Towards a single-phase mixed formulation of refractory castables and structural concrete at high temperatures

M. H. Moreira[a,b,*], R. F. Ausas[c], S. Dal Pont[d], P. I. Pelissari[a], A. P. Luz[a,b], V. C. Pandolfelli[a,b]

[a]Federal University of Sao Carlos, Graduate Program in Materials Science and Engineering (PPGCEM), Rod. Washington Luiz, km 235, 13565-905, São Carlos, SP, Brazil

[b]Federal University of Sao Carlos, Materials Microstructural Engineering Group (FIRE Associate Laboratory), Rod. Washington Luiz, km 235, 13565-905, São Carlos, SP, Brazil

[c]Institute of Mathematical and Computer Sciences, University of São Paulo, 13566-590, São Carlos, SP, Brazil

[d]CNRS, Grenoble INP, 3SR, Université Grenoble Alpes, 38000 Grenoble, France

**Abstract**

Structural materials are broadly used in applications such as nuclear vessels, high-temperature processes, and civil construction. Usually, during their placing and lifespan, they may present free or chemically bonded liquid phases in their structure, demanding careful attention when exposed to high heating rates. Their behavior in such conditions is a challenging problem as it comprises numerous highly nonlinear properties (not easily measured via experimental tests), strongly coupled equations and unreliable experimental benchmarks. Nonetheless, such simulations are of great interest. This work aims to provide a numerical study, checking whether its solution indeed converges and yields reliable results. Additionally, as the model needs several input parameters, this work conducts a sensitivity analysis and also assesses its applicability to more complex scenarios, as such issues remain open in the literature. In order to do that, a simple model that can be easily adapted for mixed formulations and complex geometries was proposed. It was found out that when considering unidimensional models the choices regarding the interpolation of the sorption isotherms are not essential to the numerical stability of the system. Besides that, the permeability and thermal conductivity of the material are the most important parameters that affect the simulation results of pressure, temperature and evaporable water content profiles. Finally, the 2D mesoscale simulation of concrete with polymeric fibers (based on the mixed formulation of the problem) yielded results that agreed with experimental observations. Thus, the model proposed herein can provide a solid base for future works and also important insights towards simpler methodologies.

**Keywords**: Thermohygro model, FEM, Porous Materials, High Temperature, Sorption Isotherms, Refractory Castable.

# Glossary

**Formula Symbols**

| | | |
|---|---|---|
| $\mathbb{P}_h$ | Piecewise constant finite element space | |
| $\mathbb{P}_k$ | Polynomial space of order $k$ | |
| $\mathbb{R}^d$ | Euclidean $d^{th}$ space | |
| $\mathcal{R}$ | Residual | |
| $\mathcal{T}_h$ | Finite element partition of the domain | |
| $\mathcal{W}_i$ | Mass content of phase i per unit volume of concrete | kg/m³ |
| $\vec{n}$ | Unit normal vector | - |
| $\vec{q}$ | Heat Flux | W/m² |
| $\vec{u}$ | Mass flux | Kg/(s m²) |
| $\vec{v}$ | Velocity | m/s |
| $C_a$ | Evaporation heat of water | J/kg |
| $C_p$ | Specific heat | J/(kg °C) |
| $D_d$ | Dufour coefficient | Pa |
| $D_s$ | Soret coefficient | T |
| $e_i$ | Error norm of variable i | |
| $g$ | Gravity acceleration | m/s² |
| $h$ | Mesh parameter | m |
| $K$ | Hydraulic conductivity | m/s |
| $L$ | Size of the 1D domain | m |
| $P$ | Pressure | Pa |
| $R$ | Ideal Gas Constant | J/(mol K) |
| $RT_k$ | Raviart-Thomas vector space of order k | |
| $T$ | Temperature | K |
| $t$ | Time | s |
| $v$ | Test function of the finite element space | |
| $w_0$ | Mass content of saturation water at room temperature per m³ of concrete | kg/m³ |
| $w_c$ | Mass content of anhydrous cement per m³ of concrete | kg/m³ |
| $X$ | Space of finite element functions | |

**Subscripts**

| | |
|---|---|
| ∞ | Ambient quantity |
| 0 | Initial content |
| an | Analytical Derivative of sorption isotherm |
| c | Cement |
| d | Dehydration |
| e | Evaporable |
| env | Environment |
| H1 | H1-Norm |
| L2 | L2-Norm |
| m | Relative to mass transfer |
| num | Numerical Derivative of sorption isotherm |
| sm | Smoothed transition of sorption isotherm |
| t | Transition |

**Other symbols**

| | | |
|---|---|---|
| $\alpha_i$ | Numerical parameter of gradient of variable i | |
| $\beta_p$ | Mass transfer coefficient | s/m |
| $\beta_T$ | Heat transfer coefficient | W/(m² °C) |
| $\delta_i$ | Differential of variable i | |
| $\Delta H_i$ | Enthalpy of phase transition of i | J/Kg |
| $\Delta t$ | Time step | s |
| $\epsilon$ | Concrete Emissivity | - |
| $\Gamma$ | Domain boundary | |
| $\kappa$ | Intrinsic permeability | m² |
| $\lambda_i$ | Thermal conductivity of phase i | W/(m K) |
| $\mu$ | Dynamic Viscosity | Pa s |
| $\Omega$ | Domain | |
| $\Phi$ | Free water content per unit volume of concrete | kg/m³ |
| $\phi$ | Relative humidity | - |
| $\psi$ | Initial Porosity | - |
| $\rho_i$ | Density of phase i | kg/m³ |
| $\sigma_{SB}$ | Stefan–Boltzmann's Constant | W/(m² K⁴) |
| $\theta$ | Numerical parameter time step | |
| $\zeta$ | Central difference numerical derivative parameter | |

# 1. Introduction

It is known that hydraulic bonded concretes can undergo thermal spalling at high temperatures due to thermal shock [1] and pressure buildup derived from the presence of water inside the porous matrix (both as physically adsorbed water and chemically-bonded comprising the hydrated phases) [2, 3, 4]. Such phenomena are especially crucial in applications that range from nuclear reactor walls [2, 5, 6], concrete buildings subject to fire [7, 8, 9, 10] to drying of refractory castable [11, 12, 13, 14, 15, 16]. Each one of these scenarios have specificities related to the context of the heating process that takes place, such as the degradation process associated with other mechanisms on the nuclear reactor walls, the fast and localized damages generated by the fire on the structural concrete or the controlled heating rate on the drying of refractory castables.

Many works presented in the literature point out that in situ measurements of some properties of structural materials are still a challenge. For instance, the placement of thermocouples and pressure sensors within the prepared samples may completely change the local properties and, thus, their overall behavior [17]. Indirect measurements may overcome such limitations [18, 19, 20], however, they are limited to small samples (i.e., cylinders of 60 mm in height and 30mm of diameter for neutron tomography analyses [18]).

Finding an alternative to estimate or predict the pressure developed inside the structures of greater dimensions is extremely valuable, as such results might be compared to the material's mechanical strength to provide guidelines for compositions that can withstand such solicitations (as required for the nuclear and civil construction applications) and to optimize industrial processes (when considering the drying of refractory castables).

Luikov et al. developed the initial studies of heat and mass transfer in capillary-porous bodies using asymptotic analysis to solve the resulting system of partial derivative equations [21], which laid the ground for further developments [2, 8, 22, 23, 9, 12, 11, 24].

With the advent of more powerful computers, numerical methods became more accessible and in the late 1970's, Bazant and colleagues developed a Finite Element Method (FEM) model that could solve a simplified version of the problem in the context of analyzing the performance of concrete applied to nuclear reactors walls [2].

Using such a model, the liquid water, air gas and water vapor were treated as a single phase. Bazant et al. reported that the convergence of the model was hard to achieve due to the transition of the sorption isotherm (which represents the state equation that depicts the amount of free water inside the material) from a non-saturated to a saturated regime and the high increase in the intrinsic permeability by two orders of magnitude.

Based on Bazant's work, Gong et al. [11] developed a FEM program that was able to estimate the pressure and temperature profiles inside a one dimensional wall of refractory castable during its drying stage. Using such a model, numerous analyses regarding which properties mostly affected the predicted pressures have been reported and which conditions favored secure and optimized drying [24, 11, 25].

Gong's model differs from Bazant's original work by the fact that it uses a simpler relationship of the sorption isotherm in the saturated region.

Whereas Bazant et al. argue that it is not realistic to consider separate flows of liquid (capillary) water, water vapor, and adsorbed water because the capillaries in hardened cement paste are not continuous [26], Tenchev et al proposed a model whose most distinguishing feature, when compared with Bazant's, was that the water vapor and liquid water were treated separately [9].

Furthermore, Davie and colleagues studied the effects of considering the capillarity effects and the mass transport of adsorbed water based on Tenchev's model [10]. Additionally, the latter model considered water diffusion which was neglected by both Bazant and Gong. Tenchev also used a smooth transition on the saturation interval of the sorption isotherm which increased the convergence of the model and was also used on further developed ones. Recently, Fey et al. [12] also applied the same approach of smoothing the sorption isotherm, but considering calculations for predicting the drying behavior of refractory castables during their first heating treatment.

In parallel under the scope of the HITECO project, Gawin et al [8] developed a thermohygromechanical model which modeled the effect of thermal and strain induced damage on construction concrete under fire. This work represented a major improvement in Bazant's model as it also considered multiple water phases, as well as the phase changes in such a system [8, 22, 23, 27].

The choices and assumptions made in the modeling procedure had a direct impact on the complexity of the resulting set of mathematical equations. Clearly, some of those choices are justified by physical and empirical reasoning, however, to the best of the author's knowledge, there is a lack of studies that analyze the influence of such decisions on the numerical behavior of the system.

The present work aims to tackle this challenge by presenting a simpler and self-containing model, which can easily be replicated. The implementation has been carried out in the open-source FEniCS Finite Element platform [28], and is one of the first publications to readily provide the resulting tool for modelling concrete and refractory castables at high temperature. FEniCS makes use of the variational formulation to declare the numerical problem through high-level programming languages such as Python and C++. This is one of the multiple advantages of FEniCS, as the basic model can be easily adapted, for instance, to use mixed formulations instead of the primal one, or define problems over more complex geometries, higher dimensions, or enhancing the modelling, for instance, considering mechanical effects, besides providing high numerical efficiency.

Considering the needs of both civil engineers and materials science engineers on having numerical models for the simulation of regular Portland cement concrete or refractory castables at high temperatures, this work also describes an unprecedented numerical convergence analysis of the model and also a sensitivity evaluation of the input parameters (which may guide one to define which property should be precisely measured and which could be obtained from the literature from similar materials). Finally, when considering a more practical case, a qualitative evaluation of the influence of adding fibers and how their presence affects the water release and pressure development during an accidentally high heating rate throughout the drying stage of a refractory castable is carried out. This was accomplished by using a novel mixed element formulation that allows to simulate systems with heterogeneous properties and high heating rates. Such analysis is also a framework for future ones on polymeric fibers design as an additive for both drying refractory castable and preventing explosive spalling for concrete under fire.

For this purpose, a 2D geometrical setting was considered, which was easily obtained from the original model in 1D, an extra benefit provided by the FEniCS platform. Finally, one of the main contributions of this work is to make the computational software available to the reader.

## 2. Mathematical model

### 2.1 Governing equations

This section presents a set of partial differential equations, the discretevariational formulation, some details of the numerical treatment adopted, and the properties selected for further analysis. This work essentially considered the mathematical formulation proposed in the seminal work by Bazant [2], with some minor adjustments to the nomenclature of the parameters (see also [24, 29] for further details).

Although Bazant's approach is based on a single-phase representation for all the fluid phases and some discussions about this aspect have already been presented in the literature [27, 26, 15], the main focus of the present study is on the development of a numerical solution of the model by using the finite element method and updated computational tools based on the FEniCS platform [28]. Besides that, some attention will be drawn to some aspects the authors have found, as well as extensive numerical experimentation.

The system of partial differential equations was derived by considering the energy and mass balance on a representative volume and the sorption isotherm curves which are the equation of state. Additionally, the unknown fields comprised temperature T, the mass water content per unit volume We, i.e., the mass of all evaporable water (not chemically bonded) and pressure p. It was also introduced in the calculations the Wd parameter as the total mass of water released by dehydration during heating, which directly depends on the temperature [2].

Consider a domain $\Omega \subset \mathbb{R}^d$; d = 1; 2 or 3 with Lipschitz boundary $\partial \Omega$. Given an initial condition We(x; 0) (the content of evaporable water) and temperature T(x; 0), the problem consists of finding T(x; t), We(x; t) and p(x; t), such that

$$\begin{cases} \rho_c \, C_{p,c} \dfrac{\partial T}{\partial t} - C_{p,w} \, \mathbf{u} \cdot \nabla T - \Delta H_e(T) \dfrac{\partial \mathcal{W}_e}{\partial t} + \nabla \cdot \mathbf{q} = -\Delta H_d \dfrac{\partial \mathcal{W}_d}{\partial t} & \text{(1a)} \\ \mathbf{q} = -\lambda_c(T) \, \nabla T & \text{(1b)} \\ \dfrac{\partial \mathcal{W}_e}{\partial t} + \nabla \cdot \mathbf{u} = \dfrac{\partial \mathcal{W}_d}{\partial t} & \text{(1c)} \\ \mathbf{u} = -\dfrac{K(T,p)}{g} \nabla p & \text{(1d)} \\ \mathcal{W}_e = \Phi(p, T) & \text{(1e)} \end{cases}$$

in which $\rho_c$ and $C_{p,c}$ are the density and isobaric specific heat of the solid matrix, $\Delta H_e$ and $\Delta H_d$ are the evaporation and dehydration enthalpies per unit mass, $\lambda_c$ is the thermal conductivity of concrete, K is the hydraulic conductivity, g the gravitational acceleration and $\Phi(p; T)$ the sorption isotherm. It was considered that the boundary of $\Omega$ was decomposed into non-overlapping parts, i.e., $\partial\Omega = \Gamma_H \cup \Gamma_C$ where the subscripts H and C stand for the hot and cold parts respectively. The system of equations presented above is subject to the following boundary conditions

$$\mathbf{u} \cdot \mathbf{n} = \beta_p (p - p_\infty) \quad \text{on } \Gamma_H \cup \Gamma_C$$
$$T = T_\infty^H(t) \quad \text{on } \Gamma_H$$
$$\mathbf{q} \cdot \mathbf{n} = \beta_T (T - T_\infty^C) \quad \text{on } \Gamma_C$$

where n is the outward unit normal, $\beta_T$ and $\beta_p$ stand for the thermal and mass film coefficients, respectively, $T_\infty^C$ the ambient temperature and $p_\infty$ is the partial pressure of the water vapor. Function $T_\infty^H$ might be defined according to the temperature description of the ISO 834 fire curve [30] or the heat up schedule defined by Gong et al. [24]. Finally, suitable initial conditions for temperature and pressure must be provided

$$T(\mathbf{x}, 0) = T_0(\mathbf{x}), \quad p(\mathbf{x}, 0) = p_0(\mathbf{x}) \tag{4}$$

This model neglects the so-called Soret effect, i.e., mass flux due to the temperature gradient because, as stated by Bazant et al [2], the thermodiffusion coefficient is small. Similarly, the Dufour effect, i.e., the heat flux due to a pressure gradient, was overlooked. Finally, the fluid velocity u is related to the Darcy velocity through

$$\mathbf{u} = \rho_w \, \mathbf{u}_{\text{Darcy}} = -\rho_w \frac{\kappa}{\mu} \nabla p \tag{5}$$

where the hydraulic conductivity K of the porous material relates to the intrinsic permeability $\kappa$ (units of m²) by

$$\kappa = K \frac{\mu}{\rho_w \, g} \tag{6}$$

## 2.2 Sorption Isotherms

The definition of the sorption isotherm $\Phi(p; T)$ deserves some attention. This is a state equation that describes the amount of free (evaporable) water contained in the material for a given temperature and relative humidity (partial pressure of water vapor). Firstly, the relative humidity is defined as

$$\phi(p, T) = \frac{p}{p_s(T)} \tag{7}$$

where $p_s(T)$ is the saturation pressure at the given temperature T. The expression for $p_s$ is given by the well-known Antoine's law which is detailed in Appendix C. The so-called sorption isotherm can be measured by fixing $\phi$ and quantifying the moisture content in the material through weight measurements [31, 2]. Bazant considered local thermodynamical equilibrium between the different water phases and deduced semi-empirical relations by correcting the theoretical results by fitting them to the experimental data [2].

As a consequence of this procedure, the hypothesis of local equilibrium does not hold and nonphysical quantities, such as relative humidity higher than one, can be found [1, 27]. It should be noted that Bazant et al. claim that the likely existence of anticlastic menisci with negative mean Gaussian curvature may justify such values of relative humidity, when considering both Kelvin and Laplace equations [26].

Gong and Mujumdar followed the same approach and adjusted the saturation behavior considering a refractory castable [24], yielding a simpler relation that is used in the present work. Finally, the sorption isotherm as a function of temperature and pressure is defined as

$$\Phi(p,T) = \begin{cases} w_c \left(\frac{w_0}{w_c} \phi(p,T)\right)^{\frac{1}{m(T)}} & \phi(p,T) \leq \phi_1 \\ \Phi_1 + (\phi(p,T) - \phi_1)\frac{(\Phi_1 - \Phi_2)}{(\phi_2 - \phi_1)} & \phi_1 < \phi(p,T) < \phi_2 \\ w_c \left[0.037(\phi - \phi_2) + 0.3335 \left(1 - \frac{T^2}{3.6\ 10^5}\right)\right] & \phi_2 \leq \phi(p,T) \end{cases} \quad (8)$$

where, $\phi_1 = 0.96$, $\phi_2 = 1.04$, $\Phi_1 = \Phi(\phi_1\ p_s(T); T)$ and $\Phi_2 = \Phi(\phi_2\ p_s(T); T)$, wc is the mass of cement per cubic meter of concrete, w0 is the saturation water quantity at ambient temperature per cubic meter of concrete, and m(T) is an empirical relation given by

$$m(T) = \phi_2 - \frac{T'}{22.34 + T'}, \quad \text{with } T' = \left(\frac{T + 10}{T_0 + 10}\right)^2 \quad (9)$$

in which the numerical constants are experimentally determined, as reported in [24].

When analyzing in more detail the expressions in Equation 8, the first part ($\phi(p;T) \leq \phi_1$) describes the behavior of an unsaturated concrete, whereas the last ($\phi_2 \leq \phi(p;T)$) is associated with a saturated one. In the interval between $\phi_1$ and $\phi_2$, Bazant proposed using a linear interpolation enforcing the values at the boundaries of such an interval to coincide with the limit values of the saturated and unsaturated regions, respectively. This procedure results in a sorption isotherm with discontinuous derivatives. As reported in [9], this may cause numerical difficulties when discretizing the problem.

A common practice is, thus, to regularize the sorption isotherm using a cubic polynomial expression for the interpolation in this transition region, so as to enforce not only the equality of the value of the sorption isotherms at the boundary of the saturation transition but also of the derivatives [9, 10, 12]. This leads to a linear system for the interpolation coefficients for each temperature that reads

$$\begin{cases} \Phi_1 = A(T)\ \phi_1^3 + B(T)\ \phi_1^2 + C(T)\ \phi_1 + D(T) \\ \frac{\partial w_1}{\partial \phi} = 3\ A(T)\ \phi_1^2 + 2\ B(T)\ \phi_1 + C(T) \\ \Phi_2 = A(T)\ \phi_2^3 + B(T)\ \phi_2^2 + C(T)\ \phi_2 + D(T) \\ \frac{\partial w_2}{\partial \phi} = 3\ A(T)\ \phi_2^2 + 2\ B(T)\ \phi_2 + C(T) \end{cases} \quad (10)$$

The results of the temperature dependent coefficients were found by using sympy, which is a symbolical algebraic Python package [32]. The final expressions for A(T), B(T), C(T) and D(T) are provided in Appendix A.

## 2.3 Intrinsic Permeability

The intrinsic permeability used here is based on the semi-empirical relationships proposed by Bazant et al [2] and Gong and colleagues [24], where a relationship between the hydraulic conductivity and the temperature and pressure is defined (knowing that the hydraulic conductivity and the intrinsic permeability are related through Equation 6). Its definition can be found in Equation 11.

$$K(P,T) = \begin{cases} K_0 f_1(\phi) f_2(T) & T \leq 368.15K \\ K_0 f_2(95) f_3(T) & T > 368.15K \end{cases} \quad (11)$$

onde,

$$f_1(p,T) = \begin{cases} \frac{1.28929 - 0.013571T}{1+[4(1-\phi(p,T))]^4} + 0.013571T - 0.28929 & \phi(p,T) < 1 \\ 1 & \phi(p,T) \geq 1 \end{cases} \quad (12)$$

The numerical constants in Equation 12 were found empirically by Gong et al. Besides that, f2(T) is defined as an Arrhenius type equation, as stated in Equation 13, describing phenomena with an associated activation energy Q = 22437J/mol and R the ideal gas constant (8.314J/(mol K)).

$$f_2(T) = \exp\left[\frac{Q}{R}\left(\frac{1}{273+T_0} - \frac{1}{273+T}\right)\right] \quad (13)$$

For temperatures higher than 95°C, the effect of f2(T) is still taken into account. However, in this case, it also depends on the temperature, as indicated by f3(T), which increases the permeability by two orders of magnitude, as a consequence of the transition between the regime dictated by the migration of water molecules along the adsorbed water layers in cement gel and the one controlled by the viscosity of the liquid water and gas mixture.

$$f_3(T) = \exp\left(\frac{T-368.15}{0.881+0.214\,(T-368.15)}\right) \quad (14)$$

## 2.4 Chemically-bonded water release

The water mass loss of castables or concrete due to dehydration is commonly obtained by using thermogravimetric analysis during the evaluation of samples previously dried at 110°C. In the current work, the chemically-bonded water release is defined as a function of the temperature considering the interpolation of the curve proposed in Gong et al. [24] as follows:

$$\mathcal{W}_d(T_k) = A_1 + (A_2 - A_1)/(1 + exp((T - T_0)/dT)) + A_3\, T \qquad (15)$$

Each coefficient value of Equation 15 is presented in Table 1.

Table 1: Coefficients used for the interpolation of Wd (Equation 15).

| $A_1$ | $A_2$ | $A_3$ | $T_0$ | $dT$ |
|---|---|---|---|---|
| 18.49 | -0.57 | 0.0073 | 267.85 | 17.34 |

## 2.5 Enthalpy of water evaporation

The enthalpy of water evaporation used in the present work is the same as the one described by Gong et al. [24]:

$$\Delta H_e(T) = \begin{cases} 3.5 \times 10^5 (374.15 - T)^{1/3} & T \leq 374.15°C \\ 0 & T > 374.15°C \end{cases} \qquad (16)$$

## 2.6 Other physical properties

The other parameters used in the model are constant and listed in Table 2. They are based on the refractory composition described by Gong et al. [24].

Table 2: Constant parameters used for the simulations, based on Gong et at. [24].

| Parameter | Value |
|---|---|
| Effective Thermal Conductivity, $\lambda_c$ | 1.67 W/(m K) |
| Density of Concrete, $\rho_c$ | 2000 Kg/m$^3$ |
| Specific Heat of Concrete, $C_{p,c}$ | 1100 J/(Kg K) |
| Specific Heat of Liquid Water, $C_{p,w}$ | 4100 J/(Kg K) |
| Enthalpy of Dehydration, $\Delta H_d$ | 0 J/Kg |
| Gravitational Acceleration, $g$ | 9.81 m/s$^2$ |
| Initial Permeability, $K_0$ | 1x10$^{-12}$ m/s |
| Cement Content, $w_c$ | 300 Kg/m$^3$ |
| Saturation Water Content at RT, $w_0$ | 100 Kg/m$^3$ |

Considering the initial and boundary conditions, in all cases the initial and environment sink temperature and pressure are given, T0 = T∞ = 298.15K and p0 = p∞ = 2850Pa. The mass and heat exchange coefficients are $\beta$T = 1W/(m²K) and $\beta$p = 1x10$^{-6}$ s/m.

Finally, the heat up curve applied to the hot face of the drying examples is defined by three different stages. Firstly, a constant heating rate of 30°C/h is applied for the initial 5.83h (up to 200°C), followed by a plateau of 10h at a constant temperature, and lastly by another step with a heating rate of 30°C/h for 14.17h (up to 625°C).

## 3. Numerical formulation

### 3.1 Preliminaries

The discretization by finite elements is straightforward for the problem (1a)-(1e). The flux variables q and u can be eliminated, T and p remain as the primary variables and a standard primal Galerkin formulation may be applied. This is the common approach adopted in the literature and the one essentially followed in this article in the convergence and sensitivity analysis. Well-posedness of this formulation is shown by Benes et al. in [29] under certain regularity assumptions. The other possibility is a mixed formulation, which will be introduced later on in the 2D case of concrete with polymeric fibers heated by the ISO curve.

Firstly, the time variable t was discretized by a subdivision of its interval into equidistant steps of size $\Delta t$. The finite element solution at $t^n = n\Delta t$ was denoted by

$$T(\mathbf{x}, t^n) \approx T_h^n, \quad p(\mathbf{x}, t^n) \approx p_h^n \qquad (17)$$

Time derivatives were approximated by finite differences

$$\left.\frac{\partial T}{\partial t}\right|_{t^{n+1}} \approx \frac{T_h^{n+1} - T_h^n}{\Delta t}, \quad \left.\frac{\partial p}{\partial t}\right|_{t^{n+1}} \approx \frac{p_h^{n+1} - p_h^n}{\Delta t} \qquad (18)$$

The temperature variable at time $n + \theta$, $\theta \in [0, 1]$ was defined by

$$T^{n+\theta} = \theta\, T^{n+1} + (1 - \theta)\, T^n \qquad (19)$$

with a similar expression for the pressure variable $p^{n+\theta}$. Moreover, for the time derivative of the water content $W_e$, it was considered:

$$\left.\frac{\partial W_e}{\partial t}\right|_{t^{n+1}} \approx \frac{\partial W_e(T^{n+\theta}, p^{n+\theta})}{\partial T}\left.\frac{\partial T}{\partial t}\right|_{t^{n+1}} + \frac{\partial W_e(T^{n+\theta}, p^{n+\theta})}{\partial p}\left.\frac{\partial p}{\partial t}\right|_{t^{n+1}} \qquad (20)$$

One of the critical points towards a robust and accurate implementation of a dryout solver is the evaluation of the sorption isotherm and its derivatives, which is pointed out in Equation (20). In the present work, the material reported by Bazant et al [2], in which the sorption is defined by a piecewise function depending on whether the concrete is saturated or not (see Equation 8 and Section 2.2), was used as a reference.

To gain further insight into the best choices for the numerical solution of the model considered for the evaluation of the castables dryout, it was analyzed the numerical behavior of several options that have been tested for the sorption isotherm function and its derivatives evaluation, namely

(a) A piecewise $C^0$ function and numerical differentiation - Equation 8 with Equation 21;

(b) The function of (a) and analytical differentiation - Equation 8 with Equations B.1 and B.2;

(c) A regularized $C^1$ function and numerical differentiation - Equation 10 with Equations A.1-A.4 and Equation 21;

(d) The function of (c) and analytical differentiation - Equation 10 with Equations A.1-A.4 with Equations B.1 and B.2;.

A second order approximation was used for the numerical differentiation, as highlighted for the derivative of We with respect to pressure:

$$\frac{\partial \mathcal{W}_e(T^n, p^n)}{\partial p} \approx \frac{\mathcal{W}_e(T^n, p^n + \delta_p) - \mathcal{W}_e(T^n, p^n - \delta_p)}{2\,\delta_p} \qquad (21)$$

where $\delta p = \zeta p^n$. Parameter $\zeta$ is taken equal to $10^{-8}$, a value chosen based on numerical experimentation. A similar expression is used for $\partial We/\partial T$. Finally, it is worth mentioning that there are other approaches in the literature that report sorption isotherms, which are not defined by piecewise functions (see [12, 22, 33]).

### 3.2 Discrete variational formulation

Let Th be a partition into finite elements of $\Omega h$, a polygonal representation of the domain, with h denoting the mesh parameter (i.e. the size for the 1D elements). The implementation was carried out with the finite element library FEniCS [28], which is a tool with enough flexibility to change from one dimensional to three dimensional problems quite straightforwardly and using triangle/quadrilateral meshes in 2D or tetrahedra/hexahedra in the 3D case.

For implementation into the FEniCS platform, the discrete variational formulation in residual form is all essentially needed. Consider the discrete finite element space defined as

$$X_h^k(\Omega_h) = \{f \in C^0(\Omega_h), f|_K \in \mathbb{P}_k(K) \;\forall K \in \mathcal{T}_h\} \qquad (22)$$

where Pk(K) is the space of polynomials of degree k on K. The problem reads:

**Problem 1.** *Given $T_h^0 = T(\mathbf{x}, 0)$ and $p_h^0 = p(\mathbf{x}, 0)$, the initial conditions, find $T_h^{n+1} \in V_{h, T_\infty(t)} = \{T_h \in X_h^{k_T}, T_h|_{\Gamma_H} = T_\infty^H(t)\}$ and $p_h^{n+1} \in W_h = X_h^{k_p}$ such that:*

$$R_T = \int_{\Omega_h} \rho\, C_{p,c} \frac{T_h^{n+1} - T_h^n}{\Delta t} v_h + \int_{\Omega_h} \lambda_c(T_h^{n+\theta}) \nabla T_h^{n+\alpha_T} \cdot \nabla v_h$$
$$+ \int_{\Omega_h} \frac{C_{p,w}}{g} K(T_h^{n+\theta}, p_h^{n+\theta}) \nabla p_h^{n+\theta} \cdot \nabla T_h^{n+\alpha_T}\, v_h -$$
$$- \int_{\Omega_h} \left( \Delta H_e \frac{\partial W_e}{\partial t} + \Delta H_d \frac{\partial W_d}{\partial t} \right)_{t^{n+1}} v_h +$$
$$\int_{\Gamma_C} \beta_T (T_h^{n+\alpha_T} - T_\infty^C) v_h = 0, \quad \forall\, v_h \in V_{h,0}(\Omega_h) \tag{23}$$

$$R_p = \int_{\Omega_h} \left( \frac{\partial W_e}{\partial t} + \frac{\partial W_d}{\partial t} \right)_{t^{n+1}} w_h$$
$$+ \frac{1}{g} \int_{\Omega_h} K(T_h^{n+\theta}, p_h^{n+\theta}) \nabla p_h^{n+\alpha_p} \cdot \nabla w_h +$$
$$+ \int_{\Gamma_H \cup \Gamma_C} \beta_p (p_h^{n+\alpha_p} - p_\infty) v_h = 0 \quad \forall\, w_h \in W_h(\Omega_h) \tag{24}$$

Regarding the adjustable numerical parameters, $\alpha_p$, $\alpha_T$, $\theta$, $k_T$ and $k_p$, it is important to highlight that:

- Linear interpolation for both variables was used in most of the numerical experiments presented in the following sections, which is equivalent to taking $k_T = k_p = 1$ for the polynomial degrees.

- Taking $\alpha_i > 0$, $i \in \{p, T\}$ was almost mandatory to avoid stringent time step restrictions due to the diffusion terms in the problem.

- The choice of $\theta$ can be critical to the nonlinear convergence and hence robustness of the model. For instance, one possibility is to linearize the problem at each time step by choosing $\theta = 0$, which works well in general and avoids iterating to find the solution at each step. Based on the authors' experience after extensive numerical assessment, the convergence of the Newton scheme, even with line search strategies, may fail depending on the problem and the heating conditions $T_\infty^H(t)$. The numerical results in this article were all obtained assuming $\theta = 0$, which was found to be the most robust choice.

- For future applications which may take advantage of increased accuracy in space and time, these numerical parameters are user-dependent on the proposed implementation and can be easily adopted.

### 3.3 Mixed formulations

One alternative to the aforementioned discrete formulation is a mixed one in which the velocity variable is retained as unknown. This can be accomplished by using a mixed strategy, leading to a three-field formulation. To that end, consider the spaces:

$$P_h(\Omega_h) = \{f \in L^2(\Omega_h),\ f|_K \in \mathbb{P}_0(K)\ \forall K \in \mathcal{T}_h\}$$
$$U_h(\Omega_h) = \{\tau \in H(div,\Omega_h),\ \tau|_K \in \mathrm{RT}_0(K)\ \forall K \in \mathcal{T}_h\}$$

where the lowest order Raviart-Thomas space is defined as

$$\mathrm{RT}_0(K) = [\mathbb{P}_0(K)]^d + \mathbf{x}\,\mathbb{P}_0(K). \tag{25}$$

Other H(div) elements can be chosen for Uh, such as a higher order Raviart-Thomas element RTk; k > 0 or the BDMk element (see Brezzi et al. [34]), which are easily handled when using the FEniCS library.

The discrete mixed formulation reads: Find $T_h^{n+1} \in V_{h,T\infty(t)} = \{Th \in X_h^{kT}, T_h|_H = T_\infty^H(t)\}$, $p_h^{n+1} \in$ Ph(Ωh) and $u_h \in$ Uh(Ωh) such that

$$\mathcal{R}_T = \int_{\Omega_h} \rho\, C_{p,c} \frac{T_h^{n+1} - T_h^n}{\Delta t} v_h + \int_{\Omega_h} \lambda_c(T_h^n) \nabla T_h^{n+1} \cdot \nabla v_h -$$
$$- \int_{\Omega_h} C_{p,w} \mathbf{u}_h^{n+1} \cdot \nabla T_h^{n+1}\, v_h -$$
$$- \int_{\Omega_h} v_h \left(\Delta H_e \frac{\partial \mathcal{W}_e}{\partial t} + \Delta H_d \frac{\partial \mathcal{W}_d}{\partial t}\right)_{t^{n+1}} +$$
$$+ \int_{\Gamma_C} \beta_T\, (T_h^{n+1} - T_\infty^C)\, v_h = 0, \quad \forall\, v_h \in V_{h,0}(\Omega_h) \tag{26}$$

$$\mathcal{R}_\mathbf{u} = \int_{\Omega_h} K^{-1}(T_h^n, p_h^n)\, g\, \mathbf{u}_h^{n+1} \cdot \boldsymbol{\tau}_h - \int_{\Omega_h} p_h^{n+1} \nabla \cdot \boldsymbol{\tau}_h +$$
$$+ \int_{\Gamma_H \cup \Gamma_C} (\beta_p^{-1} \mathbf{u}_h^{n+1} \cdot \mathbf{n} + p_\infty)\, (\boldsymbol{\tau}_h \cdot \mathbf{n}) = 0 \quad \forall\, \boldsymbol{\tau}_h \in U_h(\Omega_h) \tag{27}$$

$$\mathcal{R}_p = \int_{\Omega_h} w_h \left(\frac{\partial \mathcal{W}_e}{\partial t} + \frac{\partial \mathcal{W}_d}{\partial t}\right)_{t^{n+1}} + \int_{\Omega_h} w_h \nabla \cdot \mathbf{u}_h^{n+1} = 0 \quad \forall\, w_h \in P_h(\Omega_h) \tag{28}$$

Comparing this formulation, Equations 26-28, with the original one, Equations 23-24, some comments can be made

- The physical model and properties remain the same, the only difference being the choice of primary variables and the presence of one additional variational residual.
- The time parameters $\alpha_i$ and $\theta$ have been omitted in the unknown fields for the sake of simplicity.
- In the energy equation (Equation 26) in the third term the velocity unknown $u_h^{n+1}$ now appears.
- Note that in the second equation (Equation 27), which is the variational statement of Darcy's law, the pressure boundary condition emerges naturally in the third term.
- The potential gain of having a locally conservative formulation comes at the price of introducing an additional field, which makes the problem computationally more challenging. However, its implementation into the FEniCS platform is straightforward.

### 3.4 Software: FEniCS implementation

The FEniCS implementation enables the user to define the whole problem through the Python high-level programming language API. Due to the fact that it is easy to implement the variable parameters using Python functions and the generality of the resulting model, the decision of changing the geometry, the mesh, the boundary conditions and even switching between the primal and mixed formulations, is a simple task.

For more complex cases, one can make use of the class inheritance property of the object-orientated programming languages (such as Python and its class objects) and parameterize the whole numerical problem, such as the work described by Zhang et al [35].

Although such a strategy may be of great interest, in the current work, a simpler approach was chosen so as to avoid unnecessary abstractions. The strategy adopted was based on defining multiple flag variables (variables with capitalized names) that indicated what problem was being solved, the right set of boundary conditions, the mesh and the appropriate variational formulation. A flowchart of the FEniCS implementation as well as the piece of code used to define the variational formulation are given in Appendix D.

After considering the numerical formulation and the computational implementation in the present section, the next segment describes the numerical studies concerning the convergence analysis, the sensitivity tests and the example for the 2D mixed formulation study of adding fibers to concrete formulations, and their behavior when subjected to thermal treatment as described by the ISO 834 fire curve.

## 4. Numerical Studies

### 4.1. Convergence analysis

Convergence of the finite element formulation is one important feature to validate the implementation and assess its accuracy, especially for nonlinear problems. This is accomplished by conducting simulations with different levels of discretization in space and time.

This section aims to assess the spatial and temporal convergence rates of the proposed numerical formulation for the dryout of refractory castables or Portland cement concrete at high temperatures. In the numerical experiments below the authors report the L2 and H1-relative error norms for both pressure and temperature fields at the final time of the simulation with respect to a reference solution obtained in a very fine grid, consisting of 8000 elements and a time step $\Delta t = 1s$, i.e.,

$$e_p = \frac{\| p(t=t_f) - p_{\text{fine}}(t=t_f) \|_W}{\| p_{\text{fine}}(t=t_f) \|_W} \quad (29)$$

where W stands for $L^2(\Omega)$ or $H^1(\Omega)$. An analog expression is used for the temperature error.

The selected problem setting for this first numerical test consisted of a one-dimensional system of size L = 20 cm. The left wall TH (x = 0) was exposed to a temperature $T_\infty^H(t)$ in which two heating scenarios were considered, namely, the heat up curve described by Gong et al [24] and the ISO Fire curve [29], as presented in Equation 30. In both cases, the cold side $T_C$ was subjected to natural convection and both ends were considered permeable for the mass transfer (2), where $\beta p$ = 1.0 x 10$^{-6}$ s/m$^{-1}$ and $\beta T$ = 1W/m$^2$K.

$$T_\infty(t) = 345 \log(8t/60 + 1) + 298.15 \qquad (30)$$

Firstly, the results of the spatial convergence analysis concerning mesh refinement are given. Figures 1a, 1b, 1c and 1d show the error in the $L^2(\Omega)$ and the $H^1(\Omega)$ norms as a function of hmax for pressure (top) and temperature (bottom).

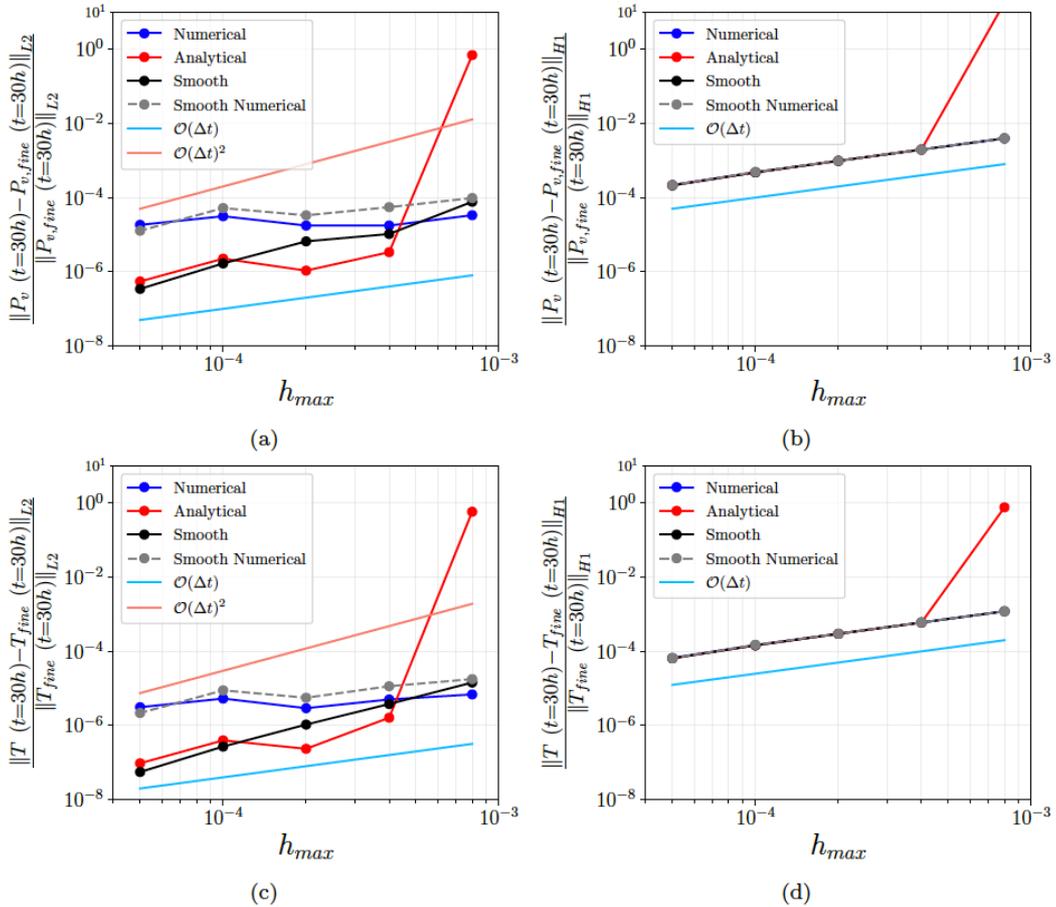

Figure 1: Mesh convergence of the pressure field, (a) and (b), and the temperature field, (c) and (e), for the drying case scenario (heat up schedule from Gong et al. [24]).

These plots point out that the applied method is convergent, however, the convergence rate depends on the specific method adopted to deal with the sorption isotherm when the $L^2(\Omega)$ norm is considered, in which case the numerical differentiation strategy has a deleterious effect. The convergence rate oscillates between ~ hmax and ~ h²max.

For the H¹-norm, although the numerical errors are significantly larger, the convergence rates behave similarly regardless of the numerical treatment considered. In all cases, the convergence rate is ~ hmax. Moving on to a fire simulation scenario, although the heating rates are higher (see Equation 30), similar trends are observed as shown in Figures 2a and 2c.

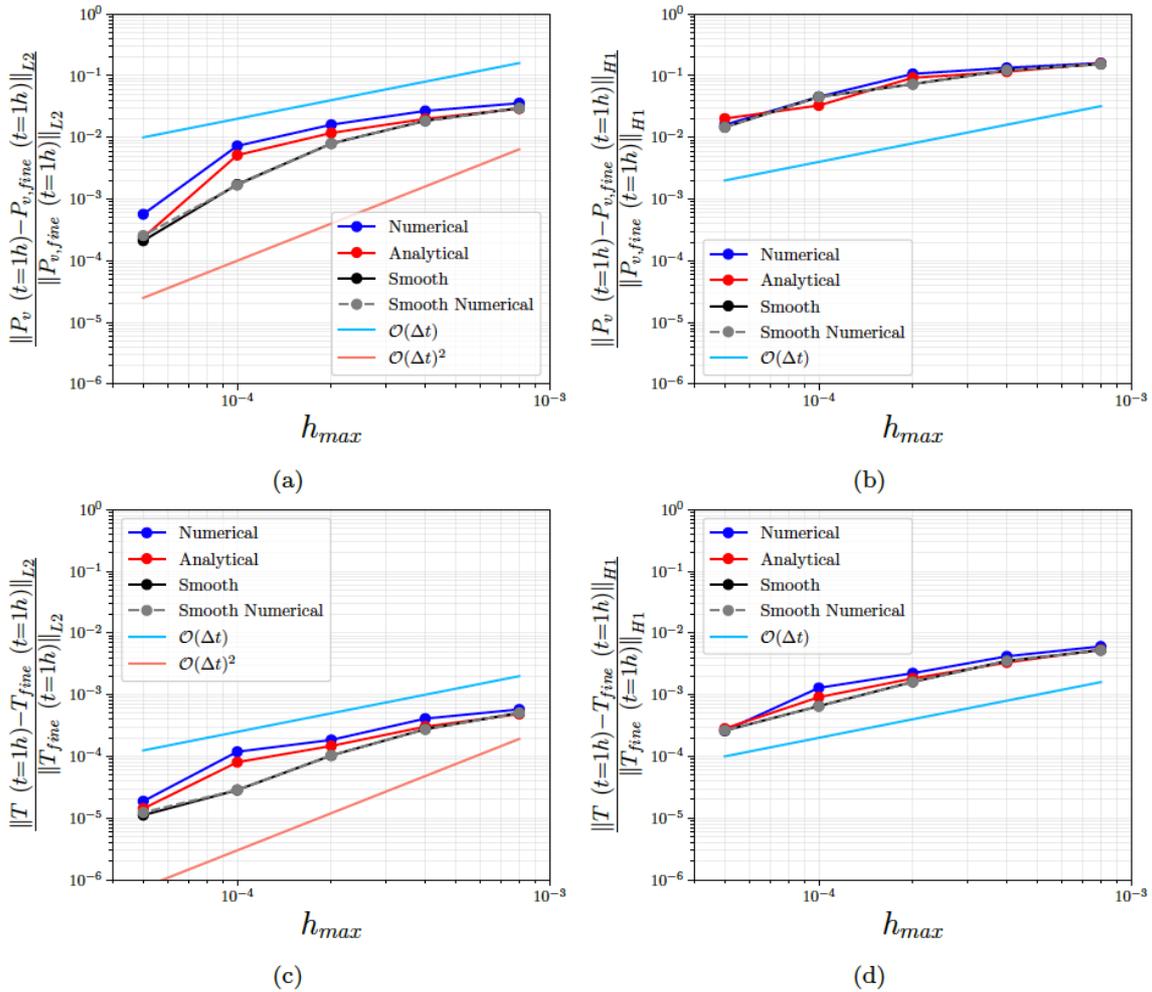

Figure 2: Mesh convergence of the pressure fields, (a) and (b), and the temperature fields, (c) and (e), for the fire case scenario (heated according to Equation 30).

However, the obtained results for the $L^2$-error norms are less sensitive to the strategy adopted for the sorption isotherm. The main conclusion is that a smooth representation of the sorption isotherm with analytical differentiation provides better results in general. Nevertheless, if ease of implementation is preferred, the other proposed treatments provide acceptable results and can be used with possible larger numerical errors.

Having assessed the spatial convergence, the next step is to evaluate the temporal convergence rate. Figure 3 shows the $L^2(\Omega)$ error norm for the pressure and temperature field as a function of the time step Δt for the two heating scenarios previously considered. The first eye-catching conclusion that can be drawn from these plots is that the method is essentially first order in time (i.e., O(Δt)). However, the asymptotic behavior is only observed for very small values of the time steps in the ISO-fire scenario, possibly due to the larger heating rates used in this case. It is worth noticing that the results seem to be quite insensitive to the choice of the sorption isotherm treatment.

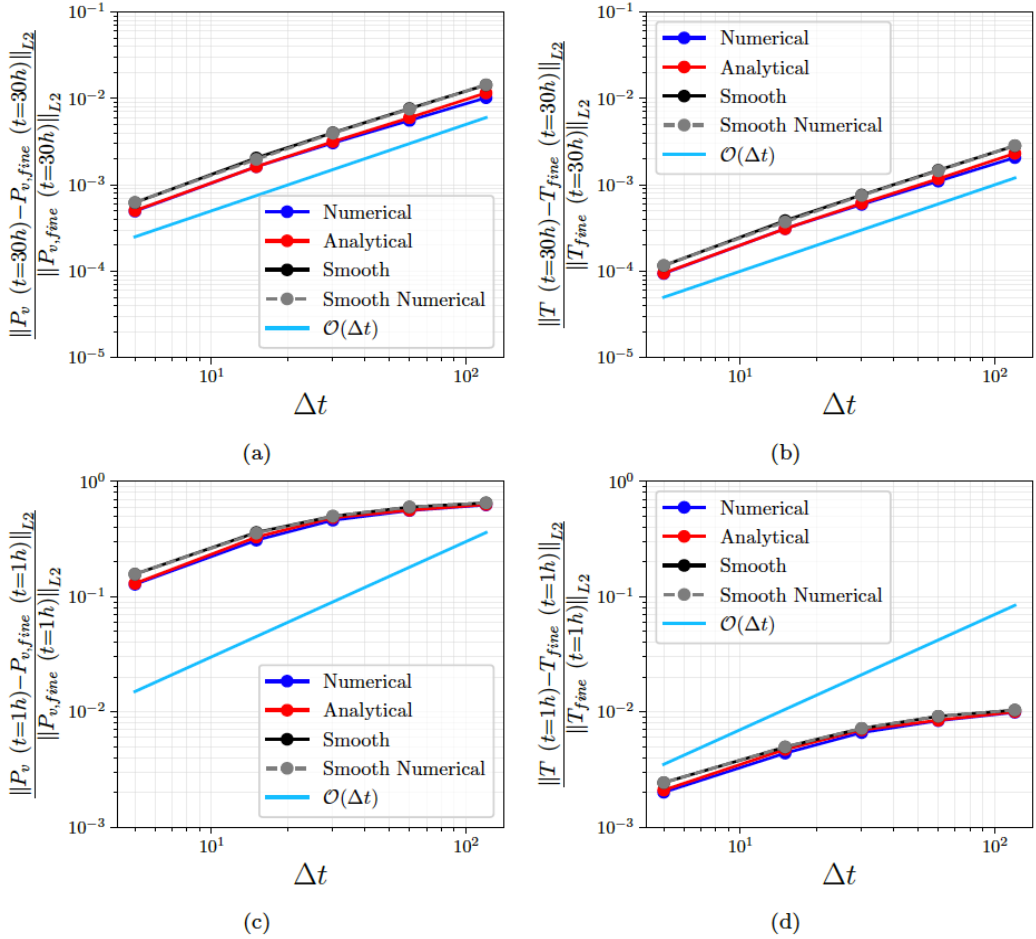

Figure 3: Time convergence considering (a-b) the drying scenario and (c-d) the fire scenario.

It can be observed that the present model displayed both convergences in time and space and that the choice of the sorption isotherm implementation does not directly affect the numerical stability of the system. Therefore, the reasoning behind the choice of the strategy may balance the ease of implementation and the accuracy of the model. Hence, the current implementation may be used for further analysis, such as the sensitivity tests and 2D analysis, which will be presented next.

**4.2 Model sensitivity for the material parameters**

The mathematical model presented above includes several physical parameters. This scenario is also quite common, and even more problematic, in multiphase and multi-component models present in different engineering applications and in particular for modeling concrete or refractory castables at high temperatures [9, 12, 8, 33]. Specifically, for the model considered here, the user must provide 10 different parameters as input to proceed with a numerical simulation. Whereas for some of these parameters there is available data in the literature or they are relatively easy to determine experimentally, others are more difficult to find out and/or they are known with high uncertainty due to inherent experimental errors.

The computational tool presented in this work is suitable to carry out a sensitivity analysis of the results for different model parameters. The authors evaluated some of the parameters that define the thermal conductivity, the intrinsic permeability and the sorption isotherm, in a range of physical feasible values and report the effect of relevant quantities, namely, the time evolution of the maximum pressure over the computational domain (which is of relative importance considering the material's damage), the temperature at the cold face and the total amount of evaporable water in the sample. The problem setting is the same as the numerical assessment of the previous section (Section 4.1).

The sensitivity of the results for the thermal conductivity was assessed by taking a specific constant value in the range between 1 and 20 W/(m K), all reasonable values that could be found in the literature for ceramic material systems [12, 24, 8]. The intrinsic permeability, on the other hand, was based on the relation described by Bazant [2] that correlates the temperature and gas pressure with the material permeability.

In such a case, the sensitivity analysis was carried out by changing the initial intrinsic permeability $K_0$ (the value measured at ambient temperature and pressure), which ranges from $10^{-14}$ m/s to $10^{-10}$ m/s. Finally, the selected sorption isotherm adopted was the one based on Bazant's model, which itself can be seen as a function parameterized by the material initial porosity and the content of anhydrous cement wc (see [2]). An additional analysis was accomplished by setting the initial porosity ranging in the interval between 3% and 15% and mass content of anhydrous cement that ranges between 165 kg/m$^3$ (7.5 wt.%) and 330 kg/m$^3$ (15 wt.%).

Figure 4a shows the maximum pressure throughout the entire simulation as a function of the initial permeability $K_0$ by fixing different values of the thermal conductivity. Two main trends were detected: (i) the decrease of the intrinsic permeability leads to an increase in the maximum pressure achieved during the simulation, which is a direct consequence of Darcy's law (see Equation 5); (ii) the increase in the thermal conductivity results in higher overall pressures, which can be explained by the fact that the amount of thermal energy conducted throughout the material is higher, providing higher temperatures for the innermost positions of the sample, and ultimately higher vapor temperatures and accordingly, higher vapor pressures.

Figure 4b displays the overall maximum pressure as a function of open porosity for different values of the mass content of anhydrous cement wc. From these plots, it can be concluded that the anhydrous cement content has a minor effect on the overall maximum pressure achieved during the simulation, whereas the initial porosity, which effectively controls the initial water content can lead to variations of up to 90% in the maximum pressure.

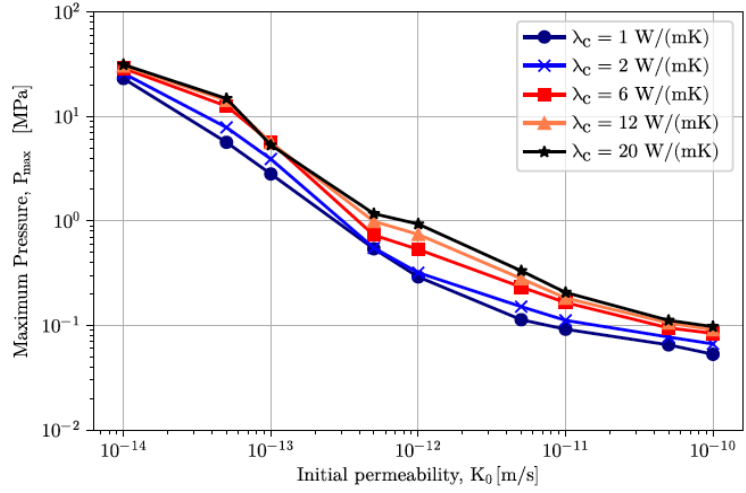

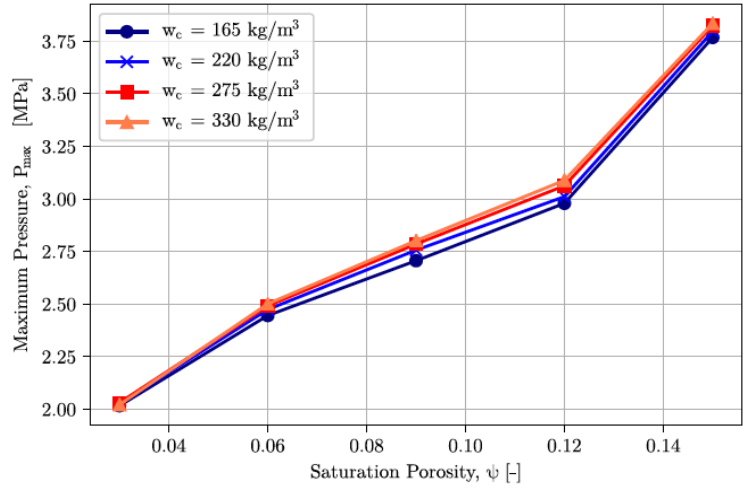

Figure 4: Overall maximum pressure throughout the entire simulation as a function of (a) intrinsic permeability for different constant thermal conductivities and (b) initial saturation porosity for different anhydrous cement content.

To gain further insight on the model behavior, a sensitivity analysis that tracks the evolution with time of the maximum pressure, the temperature on the cold side and the evaporable water content was also carried out. Figure 5 shows all these quantities as a function of time for two situations: in the left column, $K_0$ was fixed to a value of $10^{-12}$ m/s and the thermal conductivity $\lambda c$ was varied in the interval [1, 20] W/(m K), whereas, in the right column $K_0$ was considered in the range $[10^{-14}, 10^{-10}]$ m/s and the thermal conductivity had a fixed value of 4 W/(m K).

As indicated in Figure 5a, the increase in the thermal conductivity led to an overall higher maximum pressure. Moreover, the second peak, which is associated with the release of chemically bound water, presented a lower magnitude. When the material showed lower thermal conductivity, the predicted temperature evolution was slower (see Figure 5c) and, as a consequence, could be observed no clear separation between the first and the second peak in Figure 5a.

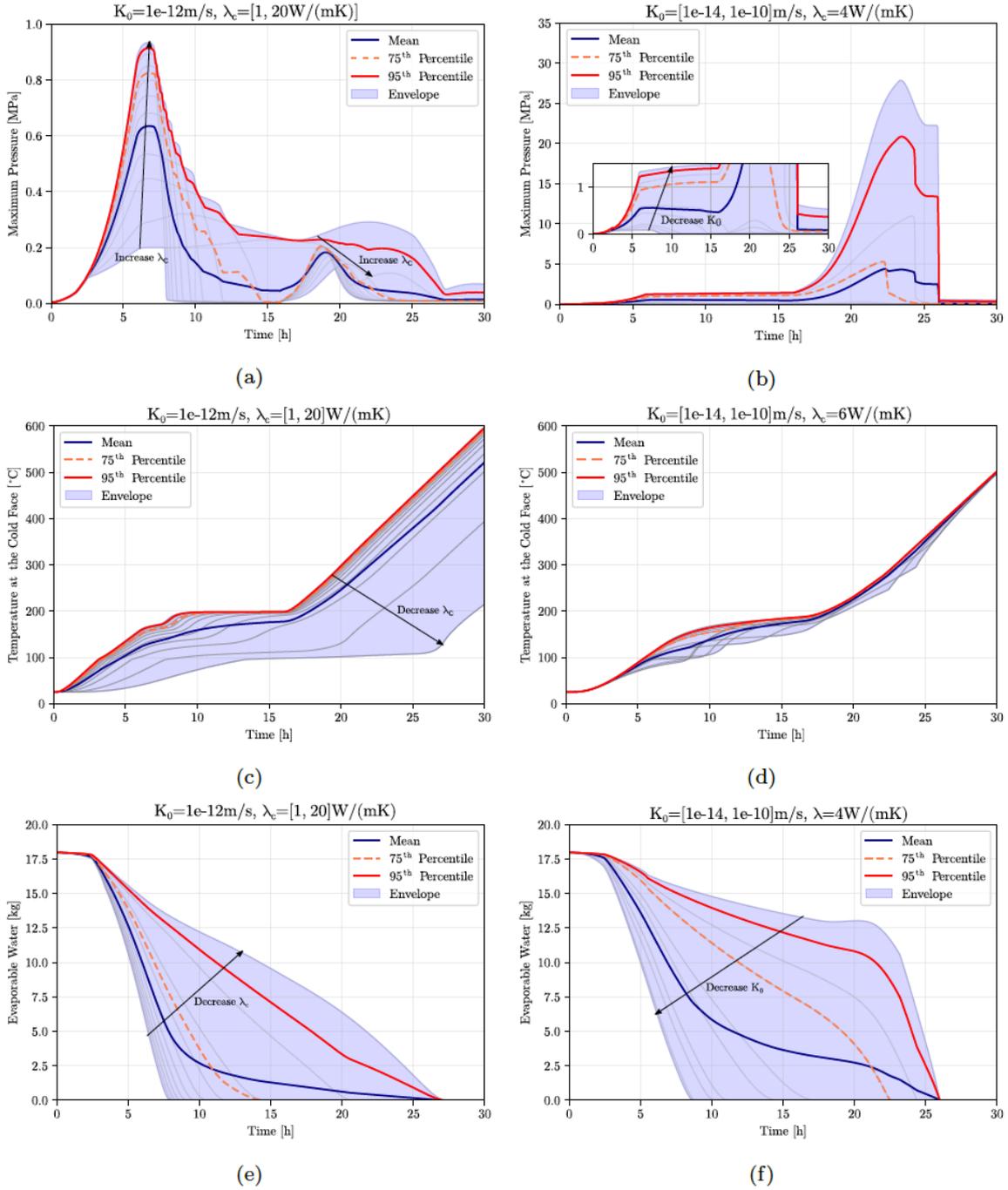

Figure 5: Maximum pressure (a-b), temperature at the cold side (c-d) and evaporable water content evolution (e-f), with fixed initial intrinsic permeability (a), (c) (e) and with fixed thermal conductivity (b), (d), (f). All the results were obtained with the heat up curve of drying described in Section 4.1.

On the other hand, when the thermal conductivity is higher, the pressure peaks also increased, leading to faster water removal and a clear separation between the peaks related to the release of free and chemically-bonded water.

Figure 5b shows that the effect of the permeability decrease can be significant. For permeabilities lower than $10^{-13}$ m/s, the second peak becomes predominant at later times, after a plateau on the maximum pressure, which can be explained by the limitation of the mass flux due to the low permeability.

Figures 5c and 5d show the effect on the temperature evolution on the cold side. From these plots, one can notice that the thermal conductivity has an important effect on the results, yielding a profile that is reminiscent of the heating up curve imposed on the hot face of the material. On the other hand, the effect of the permeability is considerably lower, and it is mostly limited for the 5-25h range.

The evaporable water content evolution is shown in Figures 5e and 5f. It can be observed that the lower the permeability, the lower the rate of water removal until the appearance of the dehydration peak (the second one on the maximum pressure evolution, Figure 5a). Two regimes can be distinguished: one related to the first pressure peak (physically-bonded water release) and the other related to the quick pressurization during the dehydration of the compounds contained in the material. The effect of lower thermal conductivities is displayed by increasing the time needed to achieve drying, yielding to roughly 27.5h to dry the material with the lowest thermal conductivity, compared to around 8h for the test with the highest $\lambda c$.

These results revealed that the initial intrinsic permeability was, indeed, the input parameter with the highest impact on the results of pressure and evaporable water content, followed by the thermal conductivity and the choice of the sorption isotherm.

Hence, methods to obtain the in situ intrinsic permeability during the phase conversions and microstructural changes that occur during the drying process are of great importance. These insights provide some guidelines which are relevant for technological applications as they indicate which model parameters are of relative importance towards an accurate prediction of concrete structures at high temperature and/or the dryout of refractory castables, for which the computational tool being proposed here suits well.

**4.3 Qualitative analysis of polymeric fibers as permeability enhancing additive**

The last set of results discussed in the present work aims to highlight the potential of the developed model based on the open source nature of the FEM solver. In real world scenarios, multiple strategies to increase the resistance of concrete structures (or refractory castable linings) exposed to high temperatures may be adopted.

Thus, adding a small amount of polymeric fibers to castable (or concrete) compositions may induce the generation of paths in the microstructure for the moisture percolation during heating, as such additives will melt and decompose. As a consequence, an increase in the overall permeability of the material should be observed, which helps to reduce the pore pressurization and the spalling risk [36, 37].

Moreover, as was observed in the last section, the permeability is the parameter that mostly affects the predicted pressure values. Thus, the following example aims to propose a qualitative analysis of using polymeric fibers as a permeability enhancing additive considering the incorporation of long and

short fibers into the refractory compositions. To simulate this case, the following assumptions were made:

- A two dimensional domain permeable in all sides (Figure 6) was selected to simulate a refractory piece heated from a single side
- The polymeric fibers were considered as subdomains of the two-dimensional refractory domain, with a constant permeability taken to be six orders of magnitude higher than the castable's initial permeability (both for the subdomains representing the short and long fibers, $\Omega 2$), representing the case where the ceramic piece has already been fired, giving rise to porous channels that increase the overall permeability. The permeability of the refractory castable domain ($\Omega 1$) was considered the sameas in the case with no fibers.
- Intersections between the fibers are likely because they were already burnt
- The thermal properties of the fibers were assumed to be equal to the properties of dry air ($\lambda f =$ 0.0262 W/(mK), $C_{p,f}$ = 1006 J/(Kg K) and $\rho f$ = 1:2754 Kg/m³)
- The heating procedure applied to the hot face follows the ISO 834 fire curve. Although the material being modeled is a refractory castable, this scenario was chosen to reproduce an extreme situation such as a failure of the heating controller. This also justified the need for a mixed formulation (slower heating rates could be simulated with the primal implementation) and make the effect of the fibers clearer.

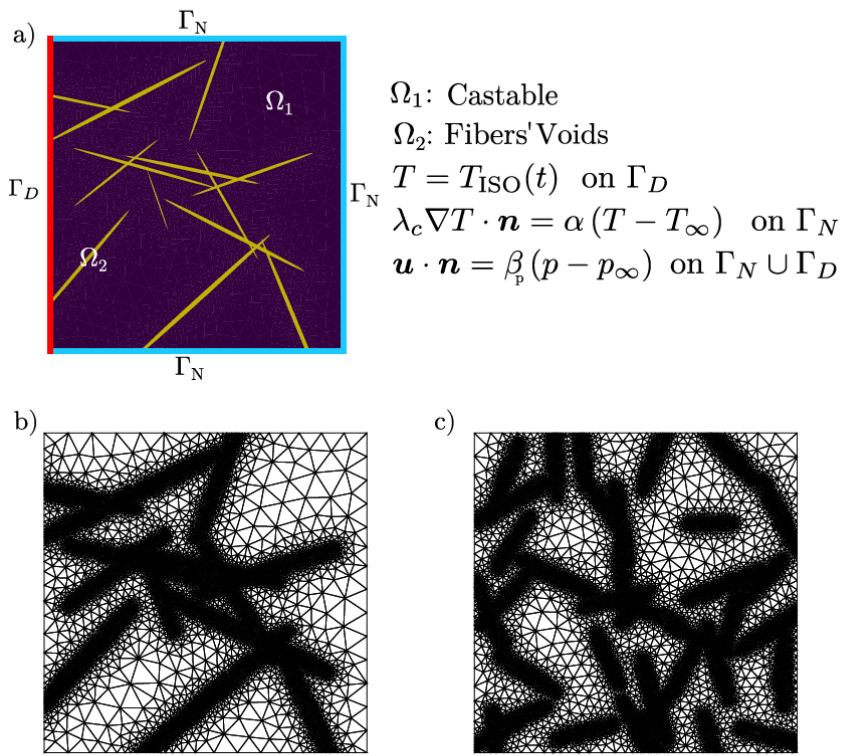

Figure 6: Description of the numerical case used to simulate and compare the effect of long and short polymeric fibers contained in a refractory castable during heating treatment (a), meshes used for the long and short fibers case, (b) and (c), respectively.

It should be noted that for further quantitative analysis, numerous aspects need to be taken into account such as the phase transformations of the polymeric fibers, the accurate value of the permeability of the regions close to them and even identify whether the validity of Darcy's law is respected in such cases.

For the generation of the domain, a simple algorithm was prepared to distribute the elliptic subdomains, $\Omega_2$, over the whole domain, $\Omega_1$, which is a square with sides of 20cm. Two distinct categories of fibers were selected, namely (i) long fibers, which have a length of $10.5 \pm 3.5$cm and (ii) short fibers, with $5 \pm 2.5$cm The case setup is described in Figure 6 and it represents a concrete slab under a ISO 834 fire curve [29, 30].

This heating procedure was also selected as the temperature increase rate is extremely high and in such cases, previous studies carried out by the authors, showed that the primal formulation (Equations 1a - 1e), was prone to instabilities. This may be explained by the fact that the primal formulation was not locally conservative and spurious mass sinks or sources might emerge in the solutions [38]. On the other hand, when using the mixed formulation proposed in this work (which has no precedent for this kind of application, to the best of the authors' knowledge) the model worked even with such high heating rates. Hence the adopted strategy is described in Equations 26 - 28.

The numerical experiments were carried out with time-steps $\Delta t = 1$s all along the total simulation time of 1h. Figure 7 describes the color maps of temperature, pressure and evaporable water content (the values were calculated using the sorption isotherm, Equation 8).

Figures 7 (a) and (b) show that the fiber positions acted as barriers for the thermal energy transport, showing regions with lower temperatures right after them. This may be explained by the fact that the regions belonging to the sub-domain $\Omega 2$ had a smaller thermal conductivity (the value used was the thermal conductivity of air) than the castable material. This also explains why the cold face stays at lower temperatures even after one hour of intense heating. Nevertheless, the present case describes an extreme situation, as in real 3D geometries, the thermal flux would be able to go around the fibers on the transverse direction, reducing the observed effect.

When comparing the effect of short and long fiber addition, it should be noticed that the randomly generated domains have roughly the same overall porosity (that is, the ratio between the area of the fibers and the complete domain), however the short fibers were randomly oriented in a way that acted as a better thermal barrier, decreasing even further the temperatures closer to the refractory's cold face. Considering the pore pressure results, Figures 7 (c) and (d) show that the longer fibers were more efficient in decreasing the pressure values on the center of the piece. Again, it should be noted that this effect is overestimated due to the bi-dimensional nature of the simulation.

Nonetheless, such results agree with experimental observations carried out by Salomão et al., who reported that fibers with longer lengths increased to a greater extent, the overall permeability of refractory castables due to their efficiency in generating interconnections between the dense matrix region and the less packed areas close to the aggregate-matrix interface [39, 40, 41], as highlighted in Figure 8.

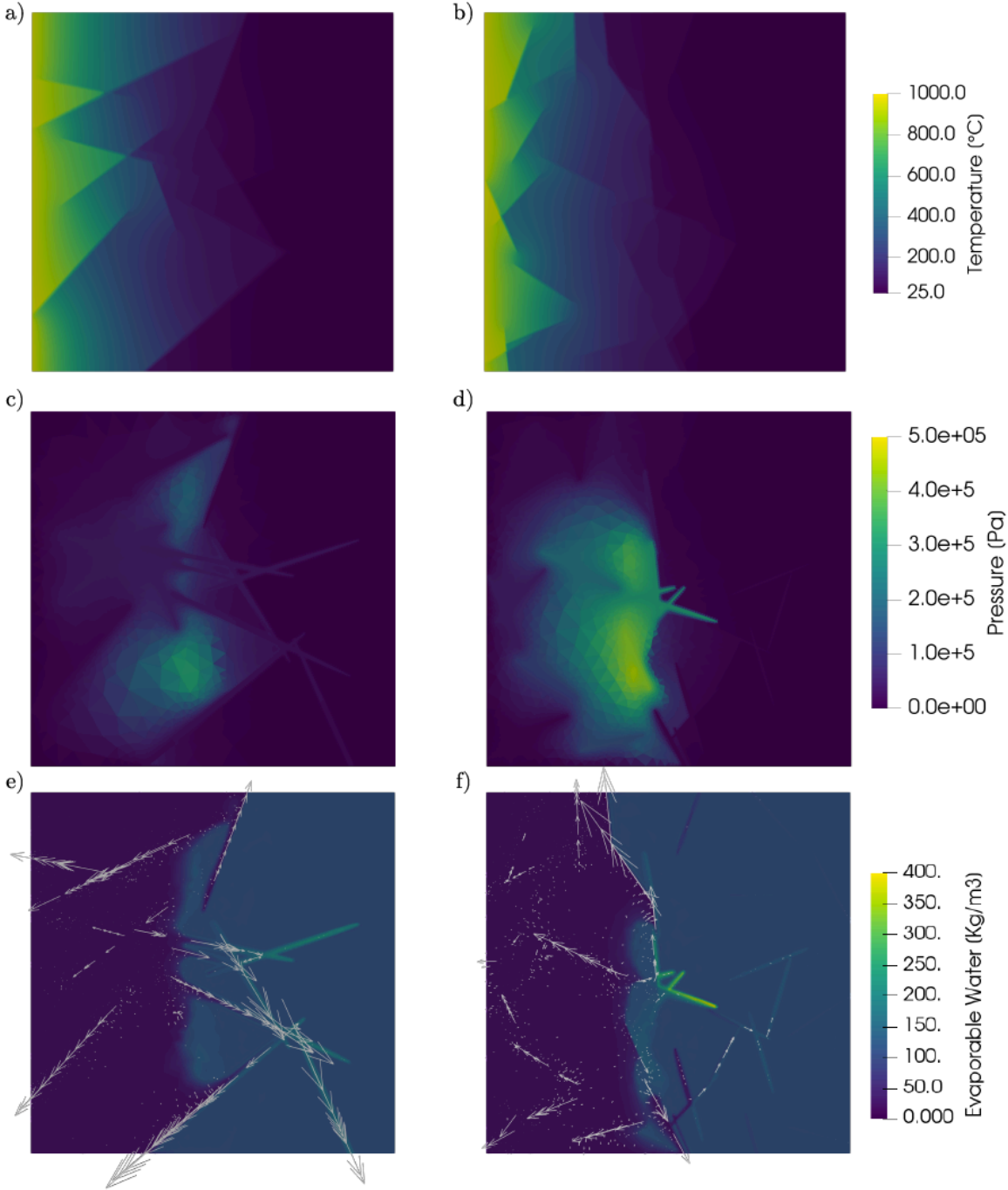

Figure 7: Results of the simulation by adding long, (a), (c) and (e), and short fibers, (b), (d) and (f), considering the temperature, (a) and (b), pressure, (c) and (d), and free water, (e) and (f), values after 1h. The vectors in (e) and (f) represent the mass flux ones, directly calculated by the mixed element formulation.

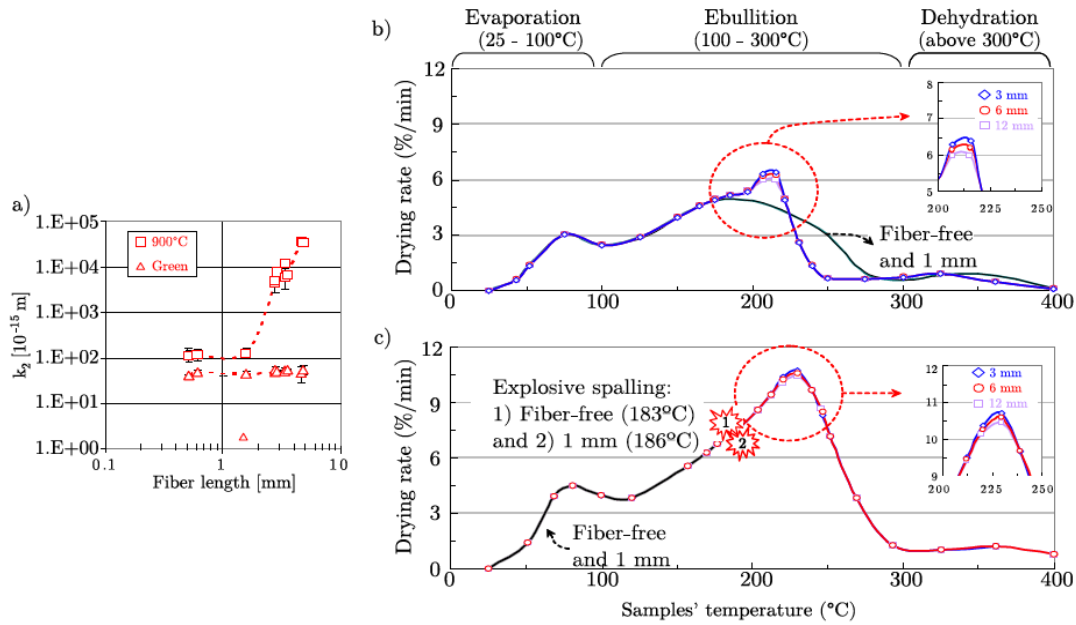

Figure 8: Experimental results of intrinsic permeability measurements with fibers of distinct sizes. (a) Forchheimer's inertial permeability constant results for refractory castables containing polypropylene fibers of different lengths for green and heated samples (900°C for six hours). Thermogravimetric tests of refractory castables with polymeric fibers with different lengths at heating rates of 10°C/min and 20°C/min, (b) and (c), respectively [39, 40].

Based on these results, fibers with longer lengths can increase the castable's permeability up to 2 orders of magnitude as they are decomposed during heating (Figure 8 (a)). This performance directly improves the explosion resistance of the samples as pointed out by the thermogravimetric tests shown in Figures 8 (b) and (c), especially when comparing the cases without fibers or with small ones (1mm) - which exploded on the TGA tests with 20°C - with those with fibers longer than 6mm - that did not explode.

Finally, when considering the simulation results for the evaporable water content after 1h of heating, Figures 7 (e) and (f), the longer fibers were also more effective in removing the water content from the castable's microstructure, which could be attained experimentally considering the higher drying rates obtained for the samples containing 12mm long polymeric fibers, Figures 8 of (b) and (c). Further analysis should be carried out in order to thoroughly confirm these findings, such as considering multiple results of randomly generated geometries of long and short fibers, or by means of representative volume analysis.

Therefore, the presented results show the potential of the developed model, despite being based on one of the oldest models previously designed for predicting the behavior of concretes at high temperatures. Its simplicity can be seen as an advantage for technological applications and even compared with more sophisticated models that demand numerous reliable measurements of complex properties, its general behavior is suitable for engineering applications [26]. Considering the extension to the mixed element formulation proposed herein, subdomains with distinct properties and extreme heating rates can be modeled, further expanding the value of this numerical tool in more complex scenarios.

5. Conclusions

The behavior of partially saturated porous construction materials at high temperatures is a subject of great importance, whether the application is the simulation of concrete civil structures under fire, or the drying of refractory castables. Given the complexity of the phenomena involved, direct and indirect experimental observations are limited. In this context, numerical simulations are highly important, providing predictions of maximum pressures and suggesting the right procedures and guidelines for projects.

In this scenario, different strategies can be made, such as the choice between multiple or single phase analyses, the methodology of managing the highly nonlinear parameters (such as the sorption isotherm), and even proposing a primal or mixed formulation.

The current work aimed to verify the numerical behavior of a single-phase simulation based on the work by Bazant et al., regarding its convergence both in time and space, which is highly important to ensure the well-posedness of the mathematical problem and assess the model's accuracy. Despite this importance, to the best of the authors' knowledge, this has not yet been reported the literature. The effect of the choices on how to implement the sorption isotherm was also analyzed. It was found that such decisions are not crucial for the numerical stability of the system when considering unidimensional models.

Once the convergence was analyzed, as the model needs several input parameters, a sensitivity analysis was proposed, pointing out that the most important input parameters for predicting the pressure developed inside the material are the intrinsic permeability of the material, its thermal conductivity and the sorption isotherms, in this order. The ease of implementation and the generality of the FEniCS platform were of great value in conducting such analysis.

A one-to-one equivalence between the resulting residual of the variational formulation to the code input was also one major benefit, especially for coupled systems of partial derivative equations such as the one analyzed herein, as this kind of simulation demands numerous input parameters which makes the implementation error-prone.

A 2D mesoscale case was also studied, considering a bi-dimensional refractory piece heated according to the procedure described in the ISO 834 fire curve and using an unprecedented mixed element formulation.

The potential of this formulation may enable the simulation of more complex cases, as the mesoscale proposed herein. The preliminary analysis of the selected example agreed with experimental observations, highlighting the good prospects of this methodology.

Thus, revisiting one of the most commonly used models for concrete and refractory systems considering its numerical behavior, provided important insights, both in theoretical and technological aspects. Most importantly, it yielded a new mixed implementation from which further advances may be discussed in further studies.

## 6. Acknowledgments

This study was financed in part by the Coordenação de Aperfeiçoamento de Pessoal de Nível Superior - Brasil (CAPES) - Finance Code 001. The authors would like to thank Conselho Nacional de Desenvolvimento Científico e Tecnologico - CNPq (grant number: 303324/2019-8) and Fundação de Amparo à Pesquisa do Estado de São Paulo - FAPESP (grant number: 2019/07996-0) for supporting this work and Túlio Mumic Cunha for the discussions and ideas. Roberto Federico Ausas also thanks FAPESP/-CEPID/CeMEAI (grant number: 2013/07375-0).

## 7. Supplementary Data

The code can be found at
https://github.com/MuriloHMoreira/supp_code_single_phase_mixed_element.